\documentclass[12pt, bezier]{article}

\def\beq{\begin{equation}}
\def\eeq{\end{equation}}
\def\beqa{\begin{eqnarray*}}
\def\eeqa{\end{eqnarray*}}

\begin{document}

\def\supp{\mathop{\rm supp}}
\def\diam{\mathop{\rm diam}}
\def\Exp{\mathop{\rm Exp}}
\def\Log{\mathop{\rm Log}}

\begin{center}

{\sc Mathematisches Forschungsinstitut Oberwolfach}
\medskip

{Report No.\ 19/2001}
\bigskip

{\large\bf Phasen\"uberg\"ange}
\medskip

{April 29th -- May 5th, 2001}
\bigskip
\bigskip
\hrule
\bigskip
\bigskip

{\large ON THE CONVERGENCE OF CLUSTER EXPANSIONS}
\medskip

{\small\sc Salvador Miracle-Sole, 
Centre de Physique Th\'eorique, Marseille}

\end{center}
\hrule 
\bigskip
\bigskip
\bigskip




\bigskip

Let ${\cal G}$ be a simple graph with a countable set ${\cal P}$ 
of vertices, also called {polymers}. 
If $\{\gamma,\gamma'\}\subset{\cal G}$ is an edge of 
the graph we say that the two polymers are
{incompatible} and write $\gamma\not\sim\gamma'$. 
Otherwise they are {compatible} and write $\gamma\sim\gamma'$. 
A complex valued function $\phi(\gamma)$,
$\gamma\in{\cal P}$, is given.
We call $\phi(\gamma)$ the {weight}, or the {activity},
of the polymer $\gamma$.
For any finite subset $\Lambda\subset{\cal P}$,
the {partition function} $Z(\Lambda)$ is defined by 
$$
Z(\Lambda)=
\sum_{{\scriptstyle S\subset\Lambda}
\atop{\scriptstyle {\rm compatible}}}
\prod_{\gamma\in S}\phi(\gamma)
$$
The sum runs over all subsets $S$ of ${\Lambda}$ such that
$\gamma\sim\gamma'$ for any two distinct elements of $S$. 
If $S$ contains only one element it is considered as a
compatible subset, and if $S=\emptyset$,
the product is interpreted as the number $1$.

A multi-index $X$ on the set ${\cal P}$ is a function 
$X(\gamma)$, $\gamma\in{\cal P}$, taking non-negative 
integer values and such that 
$\supp X$
is a finite non empty set.
Then 
$$
\ln Z(\Lambda)=
\sum_{X,\supp X\subset\Lambda}a^{\rm T}(X)
\prod_{\gamma\in{\cal P}}\phi(\gamma)^{X(\gamma)} 
=\sum_{X,\supp X\subset\Lambda}\phi^{\rm T}(X)
$$
where the coefficients $a^{\rm T}(X)$
depend only on $X$ (not on $\Lambda$), and $a^{\rm T}(X)\ne0$ 
only if $X$ is a cluster, 
i.e., only if the restriction of $\cal G$ to the vertices 
in $\sup X$ is a connected graph.

\bigskip

{\sc Theorem}: 
Assume that there is a positive function $\mu(\gamma)$, 
$\gamma\in{\cal P}$, such that, for all $\gamma_0\in{\cal P}$, 
$$
\vert\phi(\gamma_0)\vert\le\big(e^{\mu(\gamma_0)}-1\big)
\exp \bigg(-\sum_{\gamma\not\sim\gamma_0}\mu(\gamma)\bigg)
$$
Then, for all $\gamma_1\in{\cal P}$, we have
\beqa
\sum_{X,\,X(\gamma_1)\ge1}
\vert \phi^{\rm T}(X)\vert
&\le& \mu(\gamma_1)
\label{i2} \\ 
\sum_{X}
X(\gamma_1)\vert\phi^{\rm T}(X)\vert
&\le& e^{\mu(\gamma_1)}-1
\label{i3}
\eeqa
\bigskip 

The theorem is proved by an induction argument 
using the following relation:
If $\gamma_0\not\in\Lambda$, then 
$$
Z(\Lambda\cup\{\gamma_0\}) = Z(\Lambda)  
+\phi(\gamma_0)Z(\Lambda_0) 
$$
with
$\Lambda_0=\{\gamma\in\Lambda : \gamma\sim\gamma_0\}$. 

\bigskip 

More details and references can be found in 
S. Miracle-Sole,
Physica A {\bf 279}, 244--249, 2000.

\bigskip 

{\sc Note:}
To any multi-index $X$ we associate a graph
$g(X)$ with $\vert X\vert=\sum_\gamma X(\gamma)$ vertices:
$X(\gamma_i)$ distinct vertices are associated to $\gamma_i$,
for each $\gamma_i\in\supp X$. 
We draw an edge between the vertices associated
to $\gamma_i$ and $\gamma_j$ if 
$\gamma_i\not\sim\gamma_j$, and also if $\gamma_i=\gamma_j$. 
We denote by ${\cal G}(X)$ the set of all connected
subgraphs of $g(X)$ whose vertices coincide with
those of $g(X)$. 
The number of edges of a graph $g$ is denoted by $\vert g\vert$. 
If $\vert X\vert=1$ we interpret ${\cal G}(X)$ as
having one subgraph with $\vert g\vert=0$.
Then $a^{\rm T}(X)=0$ if $g(X)$ is not connected, otherwise: 
$$
a^{\rm T}(X)=
\Big(\prod_{\gamma\in{\cal P}}X(\gamma)!\Big)^{-1}
\sum_{g\in{\cal G}(X)}(-1)^{\vert g\vert} 
$$ 

This formula, and the expansion of $\ln Z(\Lambda)$, 
was proved by   
G. Gallavotti, A. Martin-Lof and S. Miracle-Sole 
in: ``Mathematical Methods in Statistical Mechanics'', 
A. Lenard, ed., pp.\ 162--202, Springer, Berlin, 1973.

\end{document}